\documentclass[10pt,notitlepage,preprintnumbers,nofootinbib,showpacs]{revtex4-1}
\usepackage{color,graphicx} 
\usepackage{ifpdf}
\usepackage{amsmath} 
\usepackage{amssymb} 
\usepackage{bm}

\usepackage{hhline}
\usepackage{color}
\definecolor{nicered}{rgb}{0.7,0.1,0.1}
\definecolor{nicegreen}{rgb}{0.1,0.5,0.1}

\usepackage[english]{babel}
\usepackage{slashed}
\usepackage{multirow}

\pdfoutput=1
\usepackage{braket}
\usepackage{slashed}

\newcommand {\E}[1]{\times 10^{#1}}	
\newcommand {\e}[1]{\mathrm{~#1}}       
\newcommand{\mc}[1]{\mathcal{#1}}

\newcommand{\mrm}[1]{\mathrm{#1}}
\newcommand{\re}[0]{\mrm{Re}}

\newcommand{\bea}{\begin{eqnarray}}
\newcommand{\eea}{\end{eqnarray}}

\newcommand{\nn}[0]{\nonumber}

\newcommand{\BR}[0]{\mathrm{BR}}

\definecolor{Red}{rgb}{1.,0.,0.}

\definecolor{Green}{rgb}{0.2,.7,0.2}

\usepackage{mciteplus}

\usepackage{hyperref}
\hypersetup{colorlinks,citecolor=nicegreen,linkcolor=nicered}
\hypersetup{colorlinks=true}

\begin{document}

\author{Svjetlana Fajfer} \email[Electronic
address:]{svjetlana.fajfer@ijs.si} 
\affiliation{Department of Physics,
  University of Ljubljana, Jadranska 19, 1000 Ljubljana, Slovenia}
\affiliation{J. Stefan Institute, Jamova 39, P. O. Box 3000, 1001
  Ljubljana, Slovenia}

\author{Nejc Ko\v snik} 
\email[Electronic address:]{nejc.kosnik@ijs.si}
\affiliation{Department of Physics,
  University of Ljubljana, Jadranska 19, 1000 Ljubljana, Slovenia}
\affiliation{J. Stefan Institute, Jamova 39, P. O. Box 3000, 1001 Ljubljana, Slovenia}

\title{Prospects of discovering new physics in rare charm decays}

\date{\today}

\begin{abstract}
  The LHCb bounds on the branching ratio of rare decay
  $D^0 \to \mu^+ \mu^-$ and the constraints on the branching ratio
  of $D^+ \to \pi^+ \mu^+ \mu^-$ in the nonresonant regions enable us
  to improve constraints on new physics contributions. Using the
  effective Lagrangian approach we determine sizes of the Wilson
  coefficients allowed by the existing LHCb bounds on rare charm
  decays. 
  Then we discuss contributions to rare charm meson decay observables
  in several models of new physics: a model with an additional spin-1
  weak triplet, leptoquark models, Two
  Higgs doublets model of type III, and a $Z'$ model. Here we
  complement the discussion by 
  $D^0 - \bar D^0$ oscillations data. Among considered models, only
  leptoquarks can significantly modify Wilson coefficients. Assuming
  that the differential decay width for $D^+ \to \pi^+ \mu^+ \mu^-$
  receives NP contribution, while the differential decay width for
  $D^+ \to \pi^+ e^+ e^-$ is Standard Model-like, we find that lepton flavor
  universality can be violated and might be observed at high
  dilepton invariant mass.
\end{abstract}

\maketitle

\section{Introduction}
Processes with charmed mesons and top quarks offer an excellent
opportunity to search for new physics (NP) in the up-type quark sector.
In contrast to B meson physics, which is convenient to search for NP due to
good exposure of the short distance effects, charm quark systems
are dominated by large long distance quantum chromodynamics
contributions. Such effects then screen
the short distance contributions of interest. Within the Standard
Model (SM) the short
distance physics in rare charm decays is strongly affected by the
Glashow-Iliopoulos-Maiani (GIM) mechanism~\cite{Glashow:1970gm}. Namely, box or penguin
diagram amplitudes get contributions from
down-type quarks which are approximately massless from the weak scale
perspective, and this warrants very effective GIM cancellation.
Flavor changing neutral current (FCNC) processes with charm mesons
might change charm quantum number for two or one unit ($|\Delta C| = 2$
or $|\Delta C|=1$ transitions). The $|\Delta C| = 2$ transition occurs in
$D^0 - \bar D^0$ oscillations and leads to strong constraints on NP
from the measured observables as pointed out in~\cite{Isidori:2010kg,Golowich:2009ii}.
There are two possibilities for NP in $|\Delta C| = 2$ transition: the
transition might occur at tree level in which case a new neutral scalar or
a vector boson possesses FCNC couplings to $u$ and $c$ quarks, or at loop level
via NP degrees of freedom affecting the box diagrams.
The processes with $|\Delta C|=1$ on the quark level are $c \to u
\gamma$ and $c \to u \ell^+ \ell^-$~\cite{Greub:1996wn,Burdman:1995te,DeBoer,Fajfer:2002gp,HoKim:1999bs,Burdman:2001tf}. 
Both transitions can be approached in the familiar effective Lagrangian formalism~\cite{Golowich:2009ii}. 
Additional constraints on NP arise from then down-type quark sector whenever
new bosons couple to left-handed quark
doublets~\cite{Dorsner:2009cu,Greljo:2015mma}. Since NP is very
constrained by the current experimental results coming from B and K
physics~\cite{Buras:2014zga} the only chance to observe NP in rare
charm decays seems to be when new bosons are coupled to weak
singlets. This then allows to avoid the strong flavor constraints in the
down-type quark sector.

On the experimental side the LHCb experiment succeeded to improve
bound on the rates of $|\Delta C| = 1$ decays by almost two orders of magnitude with respect to previous 
bounds. For the dileptonic decay the best bound to date
is~\cite{Aaij:2013cza}:
\begin{equation}
  \label{eq:Dmumu-exp}
\BR(D^0 \to \mu^+ \mu^-) < 7.6\E{-9}\,.
\end{equation}
The above limit as well as other quoted limits in the following,
unless stated otherwise,
correspond to the $95\%$ CL upper bounds. In the decay
$D^+ \to \pi^+ \mu^+ \mu^-$ the LHCb experiment focused on kinematic
regions of dilepton mass, $q^2 = (k_- + k_+)^2$, that are below or
above the dominant resonant contributions due to vector resonances in
the range $m_\rho^2 \lesssim q^2 \lesssim m_\phi^2$. The measured
total branching ratio, obtained by extrapolating spectra over the
resonant region, is~\cite{Aaij:2013sua}
\begin{equation}
  \label{eq:Dpimumu-exp}
\BR(D^+ \to \pi^+ \mu^+ \mu^-) < 8.3 \E{-8}\,,
\end{equation}
while separate branching fractions in the low- and high-
$q^2$ bins were bounded as~\cite{Aaij:2013sua}\footnote{Note that the high-$q^2$ bin quoted by
  the experiment extends beyond the maximal allowed $q^2_\mrm{max} = (m_D-m_\pi)^2 = 2.99\e{GeV}^2$.}:
\begin{equation}
  \label{eq:Dpimumu-bins}
  \begin{split}
    \BR(\pi^+ \mu^+ \mu^-)_\mrm{I} &\equiv \BR(D^+ \to \pi^+ \mu^+
      \mu^-)_{q^2 \in [0.0625,0.276]\e{GeV}^2} <  2.5\E{-8}\, \\
    \BR(\pi^+ \mu^+ \mu^-)_\mrm{II} &\equiv \BR(D^+ \to \pi^+ \mu^+
      \mu^-)_{q^2 \in [1.56,4.00]\e{GeV}^2} <  2.9
      \E{-8}\,. 
  \end{split}
\end{equation}
Motivated by these improved bounds we consider several NP models and
either derive constraints on their flavor parameters and masses, or
for the models that are severely bounded from alternative flavor
observables (e.g. $D^0-\bar D^0$ mixing, $K$, or $B$ physics), we
comment on the prospects of observing their signals in rare charm
decays.  To this end, we use the effective Lagrangian encoding
the short-distance NP contributions in a most general way.
Namely, the experimental results~\eqref{eq:Dmumu-exp} and \eqref{eq:Dpimumu-bins} 
give us a possibility to constrain NP in $c\to u \ell^+ \ell^-$ also in a
model independent way.

In the case of $b \to s \ell^+ \ell^-$ transitions, LHCb has recently
observed large departure of the experimentally determined lepton flavor universality (LFU)
ratio
$R_K= {\BR (B \to K \mu^+ \mu^-)_{q^2 \in [1,6] \rm{GeV}^2}}/{\BR (B
  \to K e^+ e^-)_{q^2 \in [1,6] \rm{GeV}^2}}$
from the expected SM value~\cite{Aaij:2014ora}. This value was found
to be $R_K^\mrm{LHCb}= 0.745^{+ 0.090}_{-0.074}\pm 0.036$, lower than the
SM prediction $R_K^{SM}= 1.0003 \pm 0.0001$~\cite{Hiller:2003js}. This
surprising result of LHCb indicates possible violation of LFU in the
$\mu$-$e$ sector. Due to the importance of this result, we investigate
whether analogous tests in the $\mu$-$e$ LFU can be carried out in $c\to u \ell^+ \ell^-$
processes.

The outline of this article is as follows.
In Section 2 we describe effective Lagrangian of
$|\Delta C|=1$ transition and determine bounds on the Wilson
coefficients coming from the experimental limits on $\BR(D^+ \to \pi^+
\mu^+ \mu^-)$ and 
$\BR(D^0 \to \mu^+ \mu^-)$.  Sec. 3 contains
analysis in the context of specific theoretical models of new physics, contributing to
the $c \to u\ell^+ \ell^-$ and related processes. Sec. 4 discusses lepton flavor
universality violation. Finally, we summarize the results and present
conclusions in Sec. 5.

\section{Observables and model independent constraints}
\subsection{Effective Hamiltonian for $c \to u \ell^+ \ell^-$}
The relevant effective Hamiltonian at scale $\mu_c \sim m_c$ is split into
three contributions corresponding to diagrams with intermediate quarks $q = d,s,b$~\cite{Burdman:2001tf,Isidori:2011qw}
\begin{align}
  \label{eq:Heff}
  {\cal H}_\mrm{eff} &= \lambda_d {\cal H}^d + \lambda_s {\cal H}^s +
                     \lambda_b {\cal H}^\mathrm{peng}\,,
\end{align}
where each of them is weighted by an appropriate combination $\lambda_q =
V_{uq} V_{cq}^*$ of Cabibbo-Kobayashi-Maskawa~(CKM) matrix elements. Virtual contributions of
states heavier than charm quark is by convention contained within
\begin{equation}
  {\cal H}^\mathrm{peng} = -\frac{4 G_F}{\sqrt{2}} \sum_{i=3,\ldots,10} C_i {\cal O}_i\,.
\end{equation}
The operators appearing in the above Hamiltonian have thus enhanced sensitivity to new physics contributions:
\begin{equation}
  \label{eq:operators}
  \begin{split}
 { \cal O}_7 &= \frac{e m_c}{(4\pi)^2}\, (\bar u \sigma_{\mu\nu} P_R c)
  \,F^{\mu\nu}\,, \qquad \quad \  {\cal O}_S = \frac{e^2}{(4\pi)^2}\, (\bar u P_R c)
  (\bar\ell  \ell)\,,\\
 {\cal O}_9 &= \frac{e^2}{(4\pi)^2}\, (\bar u \gamma^\mu P_L c)
  (\bar\ell \gamma_\mu \ell) \,, \qquad \quad {\cal O}_P = \frac{e^2}{(4\pi)^2}\, (\bar u P_R c)
  (\bar\ell \gamma_5 \ell)\, ,\\
  {\cal O}_{10} &= \frac{e^2}{(4\pi)^2}\, (\bar u \gamma^\mu P_L c)
  (\bar\ell \gamma_\mu \gamma_5 \ell)\, , \qquad   {\cal O}_T = \frac{e^2}{(4\pi)^2}\, (\bar u \sigma_{\mu\nu} c)
  (\bar\ell \sigma^{\mu\nu}\ell)\,,\\
\phantom{{\cal O}_{10} }&\phantom{= \frac{e^2}{(4\pi)^2}\, (\bar u \gamma^\mu P_L c)
  (\bar\ell \gamma_\mu \gamma_5 \ell)\, ,}
\qquad  {\cal O}_{T5} = \frac{e^2}{(4\pi)^2}\, (\bar u \sigma_{\mu\nu} c)
  (\bar\ell \sigma^{\mu\nu} \gamma_5\ell)\, .\\ 
\end{split}
\end{equation}
The chiral projectors are defined as $P_{L,R} = (1 \mp
\gamma_5)/2$, $F_{\mu \nu}$ is the electromagnetic field
strength tensor.
For each of the operators ${\cal O}_{7,9,10,S,P}$ we introduce the
corresponding counterpart ${\cal O}_{7,9,10,S,P}'$ with opposite
chiralities of quarks. Within the SM the Wilson coefficients $C_{i}$ result
from the perturbative dynamics of the electroweak interactions and QCD
renormalization. The latter effect determines the value of $C_7(m_c)$
by two-loop mixing with current-current operators and was found to be
$V_{cb}^\ast V_{ub} C_7^\mrm{SM} = V_{cs}^* V_{us} (0.007+ 0.020i) (1\pm
0.2)$~\cite{Greub:1996wn,HoKim:1999bs}.
On the other hand the value of $C_9$ Wilson coefficient was found to
be small after including renormalization group running effects as
shown in~\cite{Fajfer:2002gp} and confirmed in~\cite{DeBoer}, while
$C_{10}$ is negligible in the SM~\cite{Fajfer:2005ke}.

\subsection{$D^+ \to \pi^+ \mu^+ \mu^-$}
In order to analyze NP effects in $D^+ \to \pi^+ \mu^+ \mu^-$ one
needs to evaluate the hadronic transition matrix elements of currents
$\bar u \gamma_\mu P_{L,R} c$ and $\bar u \sigma^{\mu \nu} P_{L,R}
c$.
The standard parametrization expresses these matrix elements in terms
of three
form factors:
\begin{equation}
\Braket{\pi (k)| \bar u \gamma^\mu (1 \pm \gamma_5) c| D(p)} =
f_+(q^2)\left[(p+k)^\mu - \frac{m_D^2 - m_\pi^2}{q^2} q^\mu \right]
+ f_0 (q^2) \frac{m_D^2 - m_\pi^2}{q^2}  q^\mu \,,
\label{f1}
\end{equation}
\begin{equation}
\Braket{\pi (k)| \bar u \sigma^{\mu  \nu} (1 \pm \gamma_5) c| D(p)} = i \frac{f_T(q^2)}{m_D +m_\pi} \left[ (p+k)^\mu  q^\nu - (p+k)^\nu  q^\mu \pm i \epsilon^{\mu \nu \alpha \beta} (p+k)_\alpha   q_\beta \right]\,,
\label{f2}
\end{equation}
where $q = p-k$ is the dilepton four-momentum. For the $f_{+,0}(q^2)$
form factors we use the Be\v cirevi\' c-Kaidalov (BK) parametrization~\cite{Becirevic:1999kt}:
\begin{equation}
  \begin{split}   
f_+(q^2) &=\frac{f_+(0) }{(1-x) (1-a x)}\, ,\qquad
           x=q^2/m_\mathrm{pole}^2\,,\\
f_0(q^2) &= \frac{f_+(0)}{1-\frac{1}{b} x}\,,
  \end{split}
\label{f+}
\end{equation}
with the shape parameters $m_\mathrm{pole}$ and $a$ determined by
measurements of $D\to \pi \ell \nu$ decay spectra. We make an average
of four experimental fits to the shape parameters, by taking as input
the CLEO-c tagged~\cite{Besson:2009uv} and untagged
analysis~\cite{Dobbs:2007aa}, BES III~\cite{Ablikim:2015ixa}, and
Babar~\cite{Lees:2014ihu} results, all compiled by the
HFAG~\cite{Amhis:2014hma}.  The fitted shape parameters are
$m_\mathrm{pole} = 1.90(8)\e{GeV}$ and $a= 0.28(14)$.  For the
normalization of the form factor we rely on the lattice result
$f_+(0) =0.67(3)$ calculated by the HPQCD
collaboration~\cite{Na:2011mc}. The shape parameter $b=1.27(17)$ has
also been
extracted in lattice simulations~\cite{Fajfer:2012nr}.  For the tensor
current form factor we rely on the fit of lattice data to BK shape as
in~\cite{Fajfer:2012nr}
\begin{equation}
\label{f3a}
f_T(q^2) =\frac{f_T(0) }{(1-x) (1-a_T x)}\, ,
\end{equation}
where $x=q^2/m_{D^*}^2$, $f_T(0) =0.46(4) $ and $a_T= 0.18(16)$.
Based on the effective Hamiltonian (\ref{eq:Heff}), the most general
expression for the short distance amplitude can be written as~\cite{Bobeth:2007dw}:
 \begin{eqnarray}
 {\cal A}_\mrm{SD} (& D^+& (p) \to \pi ^+(p') \mu^+(k_+) \mu^-(k_-)) =
                       \nonumber\\
&=&\frac{iG_F  \lambda_b  \alpha}{\sqrt{2}\pi} \left[ V \, \bar u \slashed{p}
    v + A \, \bar u \slashed{p} \gamma_5 v + (S + T \cos \theta)  \bar u v+
   ( P + T_5 \cos\theta)  \bar u \gamma_5 v \right] \nn\,.
 \label{ADpi}
    \end{eqnarray}
Here $\theta$ is defined as the angle between the three-momenta of $B$ and $\ell^-$ in
the rest frame of lepton pair whereas $V,A,S,P,T$, and $T_5$ are $q^2$-dependent functions
expressed in terms of hadronic form factors and
Wilson coefficients,
\begin{eqnarray}
V &=&\frac{2m_c f_T(q^2)}{m_D+m_\pi}  (C_7 +C_7^\prime) + f_+(q^2) (C_9+C_9') +
    \frac{8 f_T(q^2) m_\ell}{m_D+m_\pi} C_T   \nn \,,\\
A &=& f_+(q^2) (C_{10} + C_{10}')  \nn\,,\\
S &=&\frac{m_D^2-m_\pi^2}{2m_c} f_0(q^2) (C_S
        +C_S')\,,\nn\\
P &=& 
      \frac{m_D^2-m_\pi^2}{2m_c} f_0(q^2)  (C_P + C_P') -m_\ell \left[f_+(q^2) -\frac{m_D^2-
      m_\pi^2}{q^2}\left(f_0(q^2)-f_+(q^2)\right)\right](C_{10}+C_{10}')\,, \nn\\
 T &=& \frac{2 f_T(q^2) \beta_\ell \lambda^{1/2}}{m_D+m_\pi} \,C_T\,,\nn\\
T_5 &=& \frac{2 f_T(q^2) \beta_\ell \lambda^{1/2}}{m_D + m_\pi} \,C_{T5}\,.
 \label{ADpi2}
 \end{eqnarray}
 We have employed introduced a shorthand notation
 $\lambda =\lambda(m_D^2, m_\pi^2, q^2)$, where
 $\lambda(x,y,z) = (x+y+z)^2-4(xy+yz+zx)$, as well as
 $\beta_\ell = \beta_\ell(q^2) = \sqrt{1-4m_\ell^2/q^2}$. The
 decay spectrum can be expressed in terms of $q^2$-dependent
 angular coefficients as:
\begin{equation}
  \frac{\mrm{d}\Gamma(D \to \pi \ell \ell)}{\mrm{d}q^2\,\mrm{d}\cos\theta} =
 N \,\lambda^{1/2} \beta_\ell \left[a_\ell(q^2) + b_\ell(q^2) \cos \theta + c_\ell(q^2)
  \cos^2\theta\right]\,,\qquad N = \frac{G_F^2 |\lambda_b|^2
\alpha^2}{(4\pi)^5 m_D^3}\,,
\end{equation}
whereas the angular coefficients are
  \begin{equation}   
  \label{eq:angcoef}
\begin{split}
  a_\ell(q^2) &=
             \frac{\lambda}{2}\left(|V|^2 +
                |A|^2 \right) + 8 m_\ell^2 m_D^2  |A|^2 +2q^2 \left[\beta_\ell^2 |S|^2 +
               |P|^2 \right]\\
  &\qquad + 4 m_\ell (m_D^2-m_\pi^2+q^2) \re[A P^*]\,,\\
\frac{b_\ell(q^2)}{4} &= q^2 \beta_\ell^2 \re[S T^*] +
              q^2 \re[P T_5^*] \\
  &\qquad+ m_\ell (m_D^2-m_\pi^2+q^2) \re[A T_5^*] +  m_\ell
              \lambda^{1/2} \beta_\ell \re[V S^*]\,,\\
c_\ell(q^2) &=  -\frac{\lambda \beta_\ell^2}{2}  \left(|V|^2 + |A|^2 \right) + 2 q^2 \left(\beta_\ell^2 |T|^2 +
              |T_5|^2 \right) + 4 m_\ell \beta_\ell\lambda^{1/2} \re[V T^*]           \,.
\end{split}
\end{equation}
The coefficients $a_\ell$ and $c_\ell$ enter then the $q^2$
distribution of branching ratio whereas $b_\ell$ is proportional to
forward-backward asymmetry:
\begin{equation}
  \begin{split}
     \frac{\mrm{d}\BR}{\mrm{d}q^2}(D \to \pi \ell \ell) &= \tau_D \,2 N
     \lambda^{1/2} \beta_\ell \left[ a_\ell(q^2) + \frac{1}{3}
       c_\ell(q^2)\right]\,,\\
     A_\mrm{FB}(q^2) &\equiv \frac{\left(\int_0^1 
         -\int_{-1}^0 \right) \mrm{d} \!\cos\theta \,\frac{\mrm{d}\Gamma(D \to \pi \ell \ell)}{\mrm{d}q^2\,\mrm{d}\cos\theta}}
{ \mrm{d}\Gamma(D \to \pi \ell \ell)/\mrm{d}q^2}
=
\frac{b_\ell(q^2)}{a_\ell(q^2) + \frac{1}{3}c_\ell(q^2)}\,.
  \end{split}
\end{equation}

\begin{figure}[t]
\centering
\begin{tabular}{cc}
\includegraphics[width=0.456\textwidth]{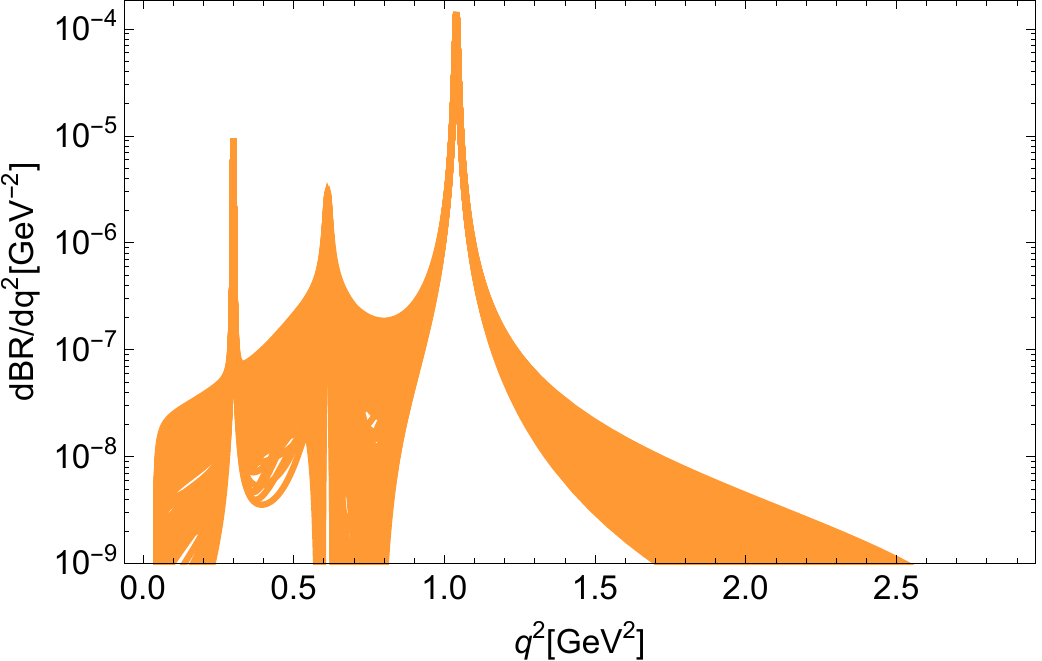}  & \includegraphics[width=0.500\textwidth]{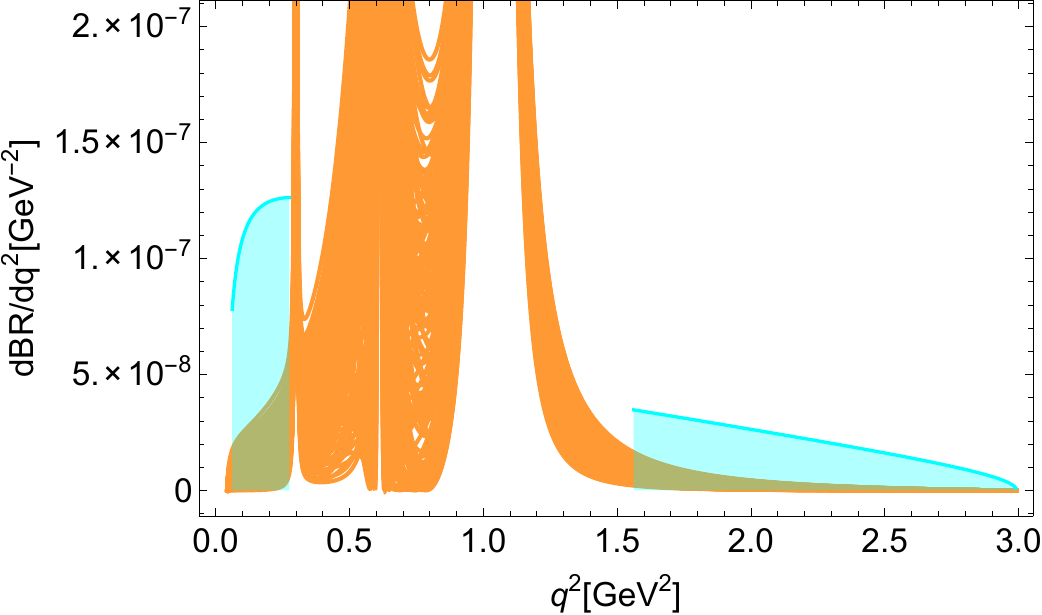}
\end{tabular}
\caption{SM resonant contributions in $D^+  \to \pi^+ \mu^+ \mu^-$
shown in orange. On the right-hand side panel cyan regions
correspond to a scenario with constant
decay amplitude that would saturate LHCb bounds~\eqref{eq:Dpimumu-bins}.}
\label{fig:LD}
\end{figure}

Contributions of the vector resonances $\rho$, $\omega$, and
$\phi$, decaying to $\mu^+ \mu^-$, is due to the first two terms in
the effective Hamiltonian~\eqref{eq:Heff} and electromagnetic
interaction. Effects of vector resonances to the
spectrum can be treated assuming na\"ive factorization by adding a
$q^2$-dependent piece to $C_9$ that contains vector
current of leptons. Analogously, the scalar contribution of $\eta$ feeds into
$C_S$. The procedure is described in detail in
Ref.~\cite{Fajfer:2007dy} for the contribution of
$D^+ \to \pi^+ \rho^0 (\omega)$ and updated for the
$D^+ \to \pi^+ \phi \to \pi^+ \mu^+ \mu^-$ in
Ref.~\cite{Fajfer:2012nr}. The current experimental upper bound
outside the resonance region indicates that the long distance
contribution is very suppressed. One might expect that at high
invariant dilepton mass bin some excited states of vector mesons might
give additional long distance contribution. However, it was shown
in~\cite{Prelovsek:2000rj} and~\cite{Fajfer:2001sa} that contributions
of these states is negligible in comparison with the leading long
distance contributions.
We parametrize the resonances with the Breit-Wigner
shapes,
\begin{equation}
\begin{split}
C_9^\mrm{res} &= \frac{\lambda_d}{\lambda_b} \left[a_\rho
  \frac{m_\rho^2}{q^2 -m_\rho^2+i \sqrt{q^2} \Gamma_\rho} + a_\omega\frac{m_\omega^2}{q^2 -m_\omega^2+i m_\omega \Gamma_\omega} - a_\phi\frac{m_\phi^2}{q^2 -m_\phi^2+i m_\phi \Gamma_\phi} \right] \,,\\
C_S^\mrm{res} &= \frac{\lambda_d}{\lambda_b} \frac{a_\eta  m_\eta^2}{q^2 -m_\eta^2+i m_\eta\Gamma_\eta}\,.
\end{split}
\end{equation}
The magnitude of unknown parameters $a_X$ ($X=\rho,\omega,\phi,\eta$),
can be fitted to the measured resonant branching ratios, given in
Tab.~\ref{tab:bw}~\cite{Agashe:2014kda}.  The corresponding values of
$|a_X|$ are given in the second row in Tab.~\ref{tab:bw}. We treat the
relative phases as free parameters.  Alternatively, for the relative
phases and magnitudes of $a_X$ one can use flavor structure
arguments~\cite{Fajfer:2005ke}. In the left-hand panel in Fig.~\ref{fig:LD} we present the
long distance contributions to the differential branching ratio for
$D^+ \to \pi^+ \mu^+ \mu^-$ as a function of dilepton invariant mass
for a representative set of parameters $|a_X|$ from the $1\sigma$
region (Tab.~\ref{tab:bw}) and random phases of $a_X$.  On the
right-hand panel in Fig.~\ref{fig:LD} we also
indicate the interpretation of experimental upper
bounds~\eqref{eq:Dpimumu-bins} in the case where the total amplitude
would be constant, namely in the case where all angular coefficient
functions $a_\ell$, $b_\ell$, $c_\ell$ would be independent of
$q^2$. We also estimate the saturation of these bounds by the total
resonant decay branching ratio and find for the low- and high-$q^2$
bin contributions to be smaller than $7.3\E{-9}$ and $5.3\E{-9}$,
respectively. On the other hand, the short distance contribution to
the total branching ratio of the SM due to the quoted value of $C_7$
is of the order $10^{-12}$ and thus negligible.
\begin{table}[!h]
  \centering
  \begin{tabular}{|c||r@{.}l|r@{.}l|r@{.}l|r@{.}l|}
    \hline
    $X$ &  \multicolumn{2}{c|}{$\rho$} &  \multicolumn{2}{c|}{$\omega$} & \multicolumn{2}{c|}{$\phi$} &  \multicolumn{2}{c|}{$\eta$}\\\hline\hline
    $\BR(D^+ \to \pi^+ X (\to\mu^+ \mu^-)) [10^{-8}]$ & $3$&$7(7)$ &
                                                                    $<3$&1
                                                                   &
                                                                         \multicolumn{1}{r}{$160$}&$\!(10)$&
                                                                                    $2$&$0(3)$\\\hline
    $|a_X|$ &$1$&$21(12)$ & $<0$&$26$ & $0$&$94(3)$ & $0$&$27(2)$\\\hline
  \end{tabular}
  \caption{$1\sigma$ ranges and $90\%$ CL upper bounds on resonant
    branching ratios and amplitude parameters~\cite{Agashe:2014kda}.}
  \label{tab:bw}
\end{table}

The branching ratio for $D^0\to \mu^+ \mu^-$ can be written in its most
general form as:
 \begin{eqnarray}
\BR(D^0 \to \mu^+ \mu^-) &=&\  \frac{1}{\Gamma_D} \frac{G_F^2 \alpha^2}{ 64 \pi^3} |V_{cb}^* V_{ub}|^2 f_D^2 m_D^3 
\beta_\mu(m_D^2)\nonumber\\
&&\times \left[  \left|\frac{2 m_\mu}{m_D} (C_{10} - C_{10}^\prime ) + \frac{m_D}{m_c} (C_P - C_P^\prime )\right|^2  
+ \frac{m_D^2}{m_c^2}\beta_\mu(m_D^2)^2\left| (C_S - C_S^\prime )\right|^2 \right]\,,
\label{BrD}
 \end{eqnarray}
where the decay constant of a $D$ meson, $f_D=
209(3)\e{MeV}$, 
has been averaged over $N_f = 2+1$ lattice
simulations~\cite{Na:2012iu,Bazavov:2011aa,FLAG}. In the SM this decay
is dominated by the intermediate $\gamma^\ast \gamma^\ast$ state that
is electromagnetically converted to a $\mu^+ \mu^-$ pair. It was estimated
in~\cite{Burdman:2001tf} that $\BR(D^0 \to \mu^+ \mu^-) \simeq
2.7\cdot 10^{-5}\times \BR(D^0
\to \gamma \gamma)$, and together with the upper bound $\BR(D^0 \to
\gamma \gamma) < 2.2\E{-6}$ at $90\%$ CL~\cite{Lees:2011qz}, this leads to
the limit $\BR(D^0 \to \mu^+ \mu^-)^\mrm{SM} \lesssim 10^{-10}$.

\section{Constraints on the Wilson coefficients}
\begin{figure}[!h]
  \centering
  \includegraphics[width=0.5\textwidth]{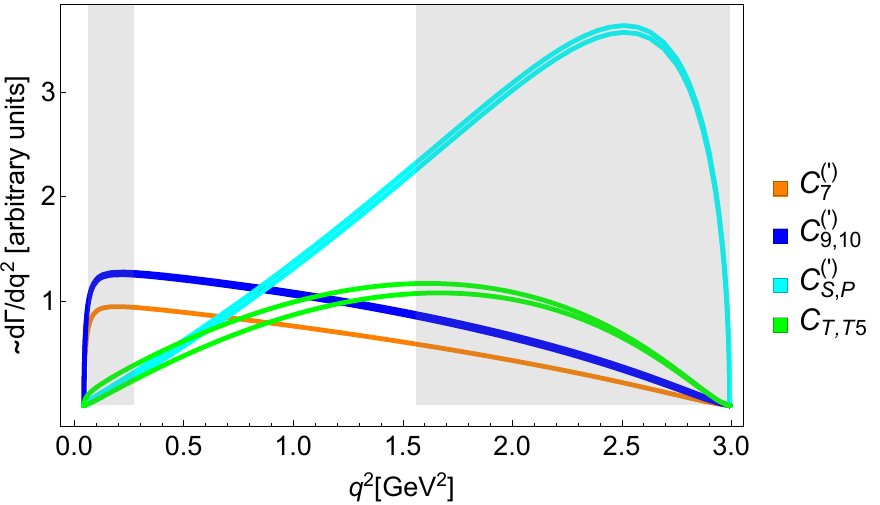}
  \caption{Comparison of short-distance spectrum sensitivities to
    different Wilson coefficients. Grey regions indicate the LHCb
    experimental low- and high-$q^2 $bins.}
  \label{fig:sens}
\end{figure}

In this Section we interpret the experimental bounds in kinematical regions I and
II given in Eq.~\eqref{eq:Dpimumu-bins} as constraints on benchmark
scenarios with NP contributions affecting individual Wilson
coefficients. In the nonresonant regions of
$D^+ \to \pi^+ \mu^+ \mu^-$ the long distance resonant contributions are one order
of magnitude below the current experimental sensitivity. This allows us to saturate experimental results for the differential decay width distribution at the low/high  dilepton invariant mass bins by the contributions of the effective Wilson coefficients. In
Fig.~\ref{fig:sens} we show the kinematical effect of setting to $1$
individual Wilson coefficients one at the time, where we have neglected the SM
resonant contributions. Strong kinematical dependence of spectra in
the cases of tensor and especially pseudoscalar coefficients suggests
they will be better constrained in the high- than in the low-$q^2$ bin. For EM dipole and (axial-)vector interactions the
enhancement at low-$q^2$ bin is hindered by relatively smaller phase space
devoted to that bin.
We allow only one Wilson coefficient at a time to have a real nonzero
value and extract its upper bound. This is repeated for each choice of
random phases and moduli of Breit-Wigner parameters
$a_{\eta,\rho,\omega,\phi}$, where the latter are sampled in their
$1\sigma$ regions (90\% CL bound for $|a_\omega|$),
c.f. Tab.~\ref{tab:bw}. The most relaxed bound obtained in this way
is then reported in Tab.~\ref{tab:BR}, where we use
notation $\tilde C_i = V_{ub} V_{cb}^* C_i$.  At the same time
the branching ratio of $D^0 \to \mu^+ \mu^-$ can give bound on Wilson coefficients $C_{10}$, $C_{S}$, and
$C_P$. It turns out that the upper bound on $\BR(D^0\to \mu^+ \mu^-)$ is
more restrictive for $C_{S,P,10}$ Wilson coefficients than any of the
invariant dilepton mass bins of $D^+ \to \pi^+ \mu^+ \mu^-$.
\begin{table}[t]
\begin{center}
\begin{tabular}{||c||r@{.}l|r@{.}l|r@{.}l||} 
\cline{2-7}
\multicolumn{1}{c||}{}&\multicolumn{6}{c||}{$|\tilde C_i|_\mathrm{max}$}\\
\cline{2-7}
 \multicolumn{1}{c||}{}  & \multicolumn{2}{c|}{$\BR(\pi \mu \mu)_\mrm{I}$} &  \multicolumn{2}{c|}{$\BR(\pi \mu \mu)_\mrm{II}$} &   \multicolumn{2}{c||}{$\BR(D^0\to \mu \mu)$} \\ 
\hline \hline
$\tilde C_7$ & \hspace{0.8cm} 2&4 \hspace{0.8cm} & \hspace{0.8cm} 1&6 \hspace{0.8cm} & \multicolumn{2}{c||}{-} \\
$\tilde C_9$ & 2&1 & 1&3 & \multicolumn{2}{c||}{-}  \\
$\tilde C_{10}$ & 1&4  & 0&92 & \hspace{.7cm}0&63\\
$\tilde C_S$ & 4&5 & 0&38& 0&049\\
$\tilde C_P$ & 3&6 & 0&37 & 0&049\\
$\tilde C_T$ & 4&1 & 0&76& \multicolumn{2}{c||}{-}  \\
$\tilde C_{T5}$ & 4&4& 0&74 & \multicolumn{2}{c||}{-}  \\
$\tilde C_9= \pm \tilde C_{10}$ & 1&3& 0&81&  0&63\\
 \hline
\end{tabular}
\end{center}
\caption{Maximal allowed values of the Wilson coefficient moduli,
  $|\tilde C_i| = |V_{ub}  V_{cb}^* C_i|$, calculated in the
  nonresonant regions of $D^+ \to \pi^+ \mu^+ \mu^-$ in the low lepton
  invariant mass region ($q^2 \in [0.0625,0.276]\e{GeV}^2$), denoted by $\mrm{I}$,
  in the high invariant mass region ($q^2 \in
  [1.56,4.00]\e{GeV}^2$), denoted by $\mrm{II}$, and from the upper
  bound $\BR (D^0 \to \mu^+ \mu^-) < 7.6 \times 10^{-9}$~\cite{Aaij:2013cza}.
  The last row gives the maximal value for the case where $\tilde
  C_9= \pm\tilde C_{10}$. All the quoted bounds have been derived for real
  $C_{i}$. The bounds for $\tilde C_i$ apply also to the chirally
  flipped coefficients $\tilde C_j^\prime$.}
\label{tab:BR}
\end{table}
The high invariant dilepton mass bin is more restrictive than low
dilepton invariant mass bin. Due to the parity conservation in $D\to
\pi$ transition the bounds for $\tilde C_j^\prime$, $ j=7,9,10,S,P$
are the same as for  $\tilde C_j$.

\begin{figure}[!h]
  \centering
  \includegraphics[width=0.5\textwidth]{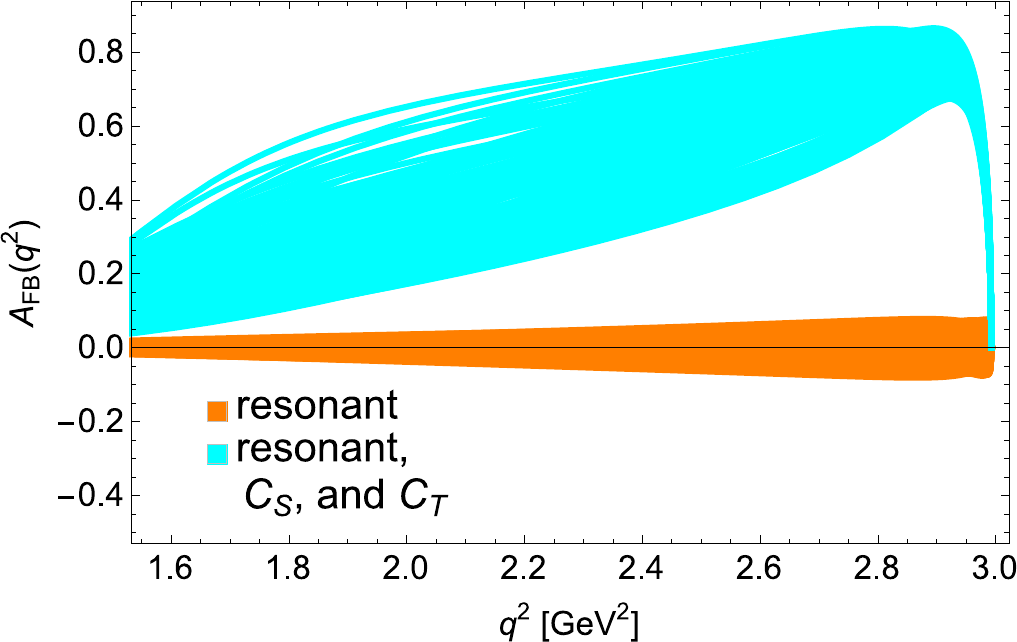}
  \caption{Forward-backward asymmetry for the resonant background
    itself (orange) and in the scenario with $C_S = 0.049/\lambda_b$,
    $C_T = 0.2/\lambda_b$ (cyan).}
  \label{fig:afb}
\end{figure}

In specific cases the angular distribution with respect to
$\cos \theta$ can be a good discriminant between the resonant and
genuine short distance contributions. It was shown that the
forward-backward asymmetry~(FBA) can be enhanced towards the larger
end of the $q^2$-spectrum in models with tensor and scalar Wilson
coefficients (or pseudoscalar and pseudotensor) simultaneously
present~\cite{Becirevic:2012fy}. In principle such a scenario can be
realized by a nonchiral leptoquark to be discussed in the following
section. As a numerical example we choose $\tilde C_S = 0.049$,
allowed by the $D^0 \to \mu^- \mu^+$, and in addition $\tilde C_T = 0.2$, which
results in $\BR(D^+ \to \pi^+ \mu^+ \mu^-)_\mrm{II} < 10^{-8}$ and is
therefore hard to distinguish from the resonant background. On the
other hand, the FBA in this case is strongly enhanced in the high-$q^2$ region, as shown in Fig.~\ref{fig:afb}.

We turn to the discussion of specific models the in next section.

\section{Impact on specific models}
\subsection{Spin-1 weak triplet}
Introducing an additional vector particle that transforms as a triplet
under $SU(2)_L$ affects a plethora of flavor observables. It
has been shown recently that a model of this type explains the current $B$
sector anomalies ($R_{D^{(*)}}$, $R_K$) even in the scenario with $U(2)_q \times U(2)_\ell$ flavor symmetry \cite{Greljo:2015mma}.
The relevant effective Lagrangian that follows from integrating out the vector
triplet at tree level reads:
\begin{equation}
{\cal L} = -\frac{1}{2m_V^2} J_\mu^a J^{a\mu}\,,
\end{equation}
representing a contact interaction between vector currents of left-handed
quark and lepton doublets:
\begin{equation}
J_\mu^a = g_q \lambda_{ij}^q \left(\bar Q^i \gamma_\mu  T^a  Q^j \right)
+  g_\ell \lambda_{ij}^\ell \left(\bar L^i \gamma_\mu T^a L^j \right)\,.
\end{equation}
The indices of $\lambda^q$ and $\lambda^\ell$ denote the
mass-eigenstates of down-type quarks and charged leptons. The
Hermitian matrices $\lambda_{ij}^{q,\ell}$ are conventionally
normalized to $\lambda_{33}^{q} = \lambda_{33}^{\ell} = 1$, with subleading entries in
$\lambda_{3i}$, whereas entries involving only the first two
generations of quarks are severely suppressed. This hierarchy is a
direct consequence of the imposed flavor symmetry.  The dominant
contributions to the processes involving only the first two
generations are induced by $\lambda_{bb}^q$ and accompanying CKM
rotations. Following \cite{Greljo:2015mma} for the rare charm decays
these are:
\begin{align}
  {\cal L}_{{\rm FCNC}} &= \frac{g_q g_\ell}{4 m_V^2} \,\lambda^\ell_{ab}  (V \lambda^q
  V^\dagger)_{ij} 
  \left(\bar \ell^a \gamma_\mu P_L \ell^b\right) \left(\bar u^i
                          \gamma_\mu P_L u^j\right)\,,
\end{align}
where the quark sector couplings originate dominantly from the CKM
mixing, namely $(V \lambda^q V^\dagger)_{uc} \approx V_{ub}
{V_{cb}}^\ast$.
Purely left-handed current effective interaction generates the
following pair of Wilson coefficients,
\begin{equation}
-C_{10} = C_9 =   R_0 \lambda^\ell_{\mu\mu} \,\frac{\pi}{\alpha}\,,
\end{equation}
where $R_0 = (g_q g_\ell m_W^2)/(g^2 m_V^2)$ is directly related to
the LFU $\tau/\ell$ ratio $R_{D^{(*)}}$ in semileptonic $B$ meson
decays whose experimental value requires $R_0 = 0.14 \pm
0.04$. Constraint from $\tau \to 3 \mu$ implies
$\lambda^\ell_{\mu\mu} = (0.013 \pm
0.011)(0.15/R_0)\frac{g_q}{g_\ell}$, while the constraint on the
$|\Delta C| = 2$ operators from CP violation in $D^0 - \bar D^0$
mixing results in an inequality $g_\ell/g_q > 1.26 R_0$.
From these ingredients one can estimate the maximum value of $C_{9}$,
\begin{equation}
 C_9  \simeq 50 \frac{0.013 \pm 0.011}{R_0}  \lesssim 10\,,
\end{equation}
which is unfortunately too small to have a detectable effect in $D \to
\pi \mu^+ \mu^-$ or in $D^0 \to \mu^+ \mu^-$.

\subsection{Leptoquarks}
There exist several scalar and vector leptoquark~(LQ) states which may
leave imprint on $c\to u \ell^+ \ell^-$ transitions~\cite{Fajfer:2008tm}. The possible
scalar states transform under the SM gauge group as $(3,3,-1/3)$,
$(3,1,-1/3)$, and $(3,2,7/6)$, of which only the latter state
conserves baryon and lepton number on the renormalizable level. Thus
the mass of the scalar multiplet $(3,2,7/6)$ can be close to the electroweak scale
without destabilizing the proton. In addition, there are four vector LQs
which potentially contribute in rare charm decays and they carry the
following quantum numbers: $(3,3,2/3)$, $(3,1,5/3)$,
$(3,2,1/6)$, and $(3,2,-5/6)$. Only the first two states have definite
baryon and lepton number.

Among all scalar LQs we will consider only the baryon number conserving state $(3,2,7/6)$
which comes with a rich set of couplings that are in general severely constrained
by $B$ and $K$ physics~\cite{Dorsner:2013tla}. Then, among the two baryon
number conserving vector LQs we will focus on the state in the
representation $(3,1,5/3)$ whose phenomenology is
limited to the up-type quarks and charged leptons.

\subsubsection{Scalar leptoquark $(3,2,7/6)$}
The renormalizable LQ couplings for the state $\Delta
(3,2,7/6)$ are~\cite{Dorsner:2013tla}
\begin{equation}
  \label{eq:LQ76}
  {\cal L} = \overline \ell_R \,Y_L\, \Delta^\dagger Q + \bar u_R \,Y_R\,
  \tilde\Delta^\dagger L + \rm{h.c.}\,.
\end{equation}
The LQ Yukawa matrices $Y_L$ and $Y_R$ are written in the mass basis of
up-type quarks and charged leptons with
the CKM and PMNS rotations present in the down-type quarks and neutrinos.
Thus, the couplings of LQ component with charge $5/3$ are
\begin{equation}
  \label{eq:Q53}
 {\cal L}^{(5/3)}= (\bar \ell_R Y_L
  u_L)\,\Delta^{(5/3)*} - (\bar u_R Y_R  \ell_L)\,\Delta^{(5/3)}  + \rm{h.c.}\,. 
\end{equation}
The tree level amplitude induced by a nonchiral LQ state $\Delta^{(5/3)}$ involves both
chiralities of fermions and is matched onto the set of
(axial)vector, (pseudo)scalar, and (pseudo)tensor operators:
\begin{equation}
  \begin{split}   
C_P = C_S &= - \frac{\pi}{2\sqrt{2} G_F \alpha \lambda_b}\, \frac{Y_{\mu
  u}^{L\ast} Y_{c \mu}^{R\ast}}{m_\Delta^2}\,,\\  
-C_P' = C_S' &= - \frac{\pi}{2\sqrt{2} G_F \alpha \lambda_b}\, \frac{Y_{\mu
  c}^{L} Y_{u \mu}^{R}}{m_\Delta^2}\,,\\ 
C_T &=-\frac{\pi}{8\sqrt{2} G_F \alpha
      \lambda_b}\,\frac{Y_{u\mu}^{R} Y_{\mu c}^{L} + Y_{c\mu}^{R*} Y_{\mu u}^{L*}}{m_\Delta^2}\,,\\
C_{T5} &= -\frac{\pi}{8\sqrt{2} G_F \alpha
      \lambda_b}\,\frac{-Y_{u\mu}^{R} Y_{\mu c}^{L} + Y_{c\mu}^{R*}
         Y_{\mu u}^{L*}}{m_\Delta^2}\,,\\
C_{10} = C_9 &= \frac{\pi}{\sqrt{2} G_F \alpha
               \lambda_b}\,\frac{Y_{\mu c}^L Y_{\mu
               u}^{L*}}{m_\Delta^2}\,\\
-C_{10}' = C_9' &= \frac{\pi}{\sqrt{2} G_F \alpha
               \lambda_b}\,\frac{Y_{c \mu}^{R*} Y_{u
                 \mu}^{R}}{m_\Delta^2}\,.
  \end{split}
\end{equation}
In the minimal numerical scenario, strict bounds in the down-type quark sector can be
evaded completely by
putting to zero the couplings to the left-handed quarks. In this case
we are allowed to have significant contributions to rare charm decays
via the $C_9^\prime = -C_{10}^\prime$ contributions for which the
bound from the last line of Tab.~\ref{tab:BR} applies. The
contribution to $D^0 - \bar D^0$ mixing amplitude is matched onto the
effective Hamiltonian $\mc{H} = C_6 (\bar u_R \gamma^\mu c_R)(\bar u_R
\gamma_\mu c_R)$ with the effective coefficient at scale $m_\Delta$
\begin{equation}
C_6(m_\Delta) = -\frac{\left(Y_{c \mu}^{R*} Y_{u \mu}^{R}\right)^2}{64\pi^2
m_\Delta^2} = -\frac{(G_F \alpha)^2}{32 \pi^4} \,m_\Delta^2 (\tilde C_{10}')^2\,.
\end{equation}
We have assumed that leptoquark does not couple to electrons or tau leptons.
Hadronic matrix element of the above operator in mixing is customarily
expressed as
$\Braket{\bar D^0 |(\bar u_R \gamma_\mu c_R) (\bar u_R \gamma^\mu c_R) |
  D^0} = \frac{2}{3} m_D^2 f_D^2 B$,
where the bag parameter in the $\overline{\mrm{MS}}$ scheme
$B_D(3\e{GeV}) = 0.757(27)(4)$ has been computed on
the lattice by the ETM Collaboration with
$2+1+1$ dynamical fermions~\cite{Carrasco:2015pra}. The SM part of the
mixing amplitude is poorly known due to its nonperturbative
nature and the only robust bound on the LQ couplings
is obtained by requirement that the mixing frequency (in the absence of
CP violation) has to be smaller than the world average $x = 2 |M_{12}|/\Gamma =
(0.49^{+0.14}_{-0.15})\%$ as quoted by the HFAG~\cite{Amhis:2014hma},
\begin{equation}
|r C_6(m_\Delta)| \frac{2 m_D f_D^2 B_D}{3 \Gamma_D}  < x\,, 
\end{equation}
where $r=0.76$ is a renormalization factor due to running of $C_6$
from scale $m_\Delta = 1\e{TeV}$ down to $3\e{GeV}$~\cite{Golowich:2009ii}.
Finally we find a bound on $C_9'$ slightly stronger but comparable
to the one obtained from $D^0 \to \mu^+ \mu^-$:
\begin{equation}
\label{eq:DDbound}
|C_6(m_\Delta)| < 2.5\E{-13}\e{GeV^{-2}} \qquad \Longrightarrow \qquad|\tilde C_{9}',\tilde C_{10}'| < 0.34\,.
\end{equation}

One can imagine an extension of this scenario which would include also
scalar and tensor operators. Namely, we consider a numerically tuned
example with $m_\Delta = 1\e{TeV}$ and large $Y_{c\mu}^R = 3$. The
bound on $C_{10}'$ from $D^0 \to \mu^+ \mu^-$ would then impose the
smallness of coupling $Y_{u\mu}^R$, $Y_{u\mu}^R < 0.007$. Bounds of
similar strength are expected from $D^0 - \bar D^0$ mixing. Now
one can introduce a nonzero coupling to left-handed quark doublet
$Y^L_{\mu u}$ that would, together with large $Y_{c\mu}^R$ contribute
to the Wilson coefficients $C_{S,P}$ and $C_{T,T5}$. However, a very
strong bound on $C_S$ now emerges from $D^0 \to \mu^+ \mu^-$ and
limits the left-handed coupling, $Y^L_{\mu u} < 1.2\E{-3}$. Thus we
can realize
\begin{equation}
  -\tilde C_{10}' = \tilde C_9' = 0.63\,,\qquad 4\tilde C_T = 4\tilde C_{T5} = \tilde C_P = \tilde C_S = -0.049\,,
\end{equation}
together with small enough $Y^L_{\mu u} = 1.2\E{-3}$
to comply with the constraints from $B$, $K$ physics and four fermion operator
constraints~\cite{Carpentier:2010ue}.

\subsubsection{Vector leptoquark $(3,1,5/3)$}
The interactions of the vector LQ state $V^{(5/3)} (3,1,5/3)$ with the SM fermions are
contained in a single term at the renormalizable level:
\begin{equation}
  \label{eq:V53}
  \mc{L} = Y_{i j}\, (\bar \ell_i \gamma_\mu P_R u_j)\, V^{(5/3)\mu} + \mrm{h.c.}\,.
\end{equation}
Generation indices are denoted by $i,j$. Integrating out $V^{(5/3)}$
results in
the right-handed current operators:
\begin{equation}
C_9' = C_{10}' = \frac{\pi}{\sqrt{2} G_F \lambda_b \alpha} \,
\frac{Y_{\mu c} Y_{\mu u}^*}{m_V^2}\,.  
\end{equation}
On the other hand, the same combination of couplings enters the $D^0-\bar D^0$
mixing. We employ the same type of Hamiltonian as in the preceding
Section this time the Wilson coefficient:
\begin{equation}
C_6(m_V)= \frac{(Y_{ \mu u}Y^*_{\mu c})^2 }{32 \pi^2 m_V^2} =
\frac{(G_F \alpha)^2}{16\pi^4} m_V^2 (\tilde C_{10}')^2\,.
\end{equation}
Consequence of the bound~\eqref{eq:DDbound} is that the rare decay
Wilson coefficients are limited:
\begin{equation}
|\tilde C_9',\tilde C_{10}'| < 0.24\,.  
\end{equation}
The above knowledge of $C_{9,10}^{\prime}$ implies that the branching ratio of $D \to \pi \mu^+ \mu^-$ in the high-$q^2$ bin 
is at most $1.4\E{-8}$, where the long-distance uncertainties have
been taken into account. The effect is twice smaller than the existing experimental bound.

\subsection{Two Higgs doublet model type III}
In the Two Higgs Doublet Model of type III (THDM III) the neutral
Higgses have flavor changing couplings to the fermions. The spectrum
includes two neutral scalars, $h$ and $H$, one pseudoscalar, $A$, and
two charged scalars, $H^\pm$. In the scenario with MSSM-like scalar potential their
masses and mixing angles are related~\cite{Crivellin:2013wna},
\begin{equation}
\begin{split}
\tan\beta &= \frac{v_{u}}{v_{d}},\quad \tan 2\alpha= \tan 2\beta\frac{m_{A}^{2}+m_{Z}^{2}}{m_{A}^{2}-m_{Z}^{2}}\,,\\
m_{H^{\pm}}^{2} &= m_{A}^{2}+m_{W}^{2}\quad m_{H}^{2} = m_{A}^{2}+m_{Z}^{2}-m_{h}^{2}\,,
\end{split}
\end{equation}
where $\beta$, $\tan\beta = v_u/v_d$, is the angle that diagonalizes
the mass matrix of the charged states, $\alpha$ is the mixing angle of
neutral scalars.
The vacuum expectation values are normalized to the electroweak vacuum
expectation value,
$v/\sqrt{2}=\sqrt{v_u^2+v_d^2} = 174$~GeV.
The part of the interaction Lagrangian responsible for FCNCs in the up-type quarks and
charged leptons is~\cite{Crivellin:2013wna}
\begin{equation}
  \begin{split}
\mathcal{L} = \frac{y_{ij}^{(\ell)H_k}}{\sqrt{2}}H_k
\bar{\ell}_{L,i}\ell_{R,j}+ \frac{y_{ij}^{(u)H_k}}{\sqrt{2}}H_k
\bar{u}_{L,i} u_{R,j}+\textrm{h.c.}\,, \qquad H_k = (H,h,A)\,,
  \end{split}
\end{equation}
and the neutral Yukawa couplings for the charged leptons and up-type quarks are
\begin{equation}
  \begin{split}
y_{ij}^{(\ell)H_{k}} &=
                       x_{d}^{k}\frac{m_{\ell_{i}}}{v_{d}}\delta_{ij}-\epsilon_{ij}^{\ell}\left(x_{d}^{k}\,\tan\beta-x_{u}^{k*}\right)
                       \,,\\
y_{ij}^{(u)H_{k}} &= x_{u}^{k}\frac{m_{u_i}}{v_u}\delta_{ij}-\epsilon_{ij}^{u}\left(x_{u}^{k}\,\cot\beta-x_{d}^{k*}\right) \,,
  \end{split}
\end{equation}
respectively. The flavor off-diagonal terms $\epsilon^\ell_{fi}$, $\epsilon^u_{fi}$ are free
parameters of the model. The coefficients $x_q^k$ for $H_k=(H,h,A)$
are determined by the mixing angles of the neutral scalars and the VEVs~\cite{Crivellin:2013wna}
\begin{equation}
\begin{split}
x_{u}^{k} &=\left(-\sin\alpha,-\cos\alpha,i\cos\beta\right)\,, \\
x_{d}^{k} &= \left(-\cos\alpha,\sin\alpha,i\sin\beta\right)\,.
\end{split}
\end{equation}
For the transitions $c \to u\ell^+ \ell^-$ the driving flavor changing
parameter is $\epsilon_{12}^u$ that induces scalar and pseudoscalar
Wilson coefficients, while we assume that $\epsilon_{22}^\ell$ is negligible~\cite{Crivellin:2013wna}:
\begin{align}
-C_P = C_S &= \frac{\pi}{4\sqrt{2} G_F \alpha \lambda_b}
\,\frac{m_\mu}{v}\frac{\epsilon_{12}^{u*} \tan\beta}{m_H^2}\,,\\
C_P' = C_S' &=  \frac{\pi}{4\sqrt{2} G_F \alpha \lambda_b}
\,\frac{m_\mu}{v}\frac{\epsilon_{21}^{u} \tan\beta}{m_H^2}\,.
\end{align}
The best upper bounds on $C_P$, $C_S$, or $C_P'$, $C_S'$ pairs are
obtained from $\BR(D^0 \to \mu^+ \mu^-)$ and read $|\tilde C_S -
\tilde C_S'| \leq 0.05$ and $|\tilde C_P - \tilde C_P^\prime| \leq 0.05$ which makes them very difficult to probe in $D \to \pi \mu^+
\mu^-$ decay, unless the cancellation between $C_S$ ($C_P$)
and $C_S'$ ($C_P'$) in $D^0 \to \mu^+ \mu^-$ is arranged by fine-tuning.

\subsection{Flavor specific $Z^\prime$ extension}
An additional neutral gauge boson appears in many extensions of the
SM. Current searches for $Z^\prime$ at the LHC are well motivated by
many extensions of the SM, see
e.g.~\cite{Celis:2015ara,Langacker:2000ju}. Even more, a $Z^\prime$
boson can explain $B\to K^* \mu^+ \mu^-$ angular asymmetries puzzle,
as presented in e.g.~\cite{Descotes-Genon:2015xqa,Buras:2012jb}. Assuming as
in~\cite{Langacker:2000ju} flavor nonuniversal couplings of
$Z^\prime$ to fermions, we allow $Z'$ to couple only to the pair $\bar c u$ and
$c \bar u$. Such model in the most general way has been considered by the
authors of~\cite{Golowich:2009ii}. In order to avoid constraints
coming from the down-type quark sector which will affect left-handed quark
couplings, we allow only right-handed couplings of
${\cal L} _{Z^\prime}^q = C^u (\bar u \gamma^\mu P_R c) Z^\prime_\mu$.
This assumption leads to the same effective operator
${\cal H}^\mrm{eff} = C_6 (\bar u \gamma_\mu P_R c) (\bar u \gamma^\mu
P_R c)$
as already discussed in the case of leptoquarks. The effective
Wilson coefficient describing $D^0 -\bar D^0$ transition is now:
\begin{equation}
C_6(m_{Z^\prime}) =\frac{|C^u |^2}{ 2 m_{Z^\prime}^2}\,.
\label{Z1}
\end{equation}
The bound on $C_6$~\eqref{eq:DDbound} leads to $|C^u|< 7.1\E{-4}  (m_{Z^\prime} /1\e{TeV})$. 
Allowing $Z^\prime$ to couple to muons as in the SM  with $g_L^\ell =
(g/\cos\theta_W )(-1/2 +\sin^2 \theta_W) $ and $g_R^\ell= g \sin^2
\theta_W/\cos\theta_W$, we obtain 
\begin{equation}
C_9^\prime =  \frac{4 \pi}{{\sqrt 2} G_F \lambda_b \alpha} \frac{(g_L^\ell + g_R^\ell) C^u }{2 m_{Z^\prime}^2}
\label{Z9}
\end{equation}
and 
\begin{equation}
C_{10}^\prime =  \frac{4 \pi}{{\sqrt 2} G_F \lambda_b \alpha} \frac{(- g_L^\ell + g_R^\ell) C^u }{2 m_{Z^\prime}^2}\, .
\label{Z10}
\end{equation}
For $m_{Z^\prime}  \sim 1$ TeV this amounts to $|C_9| \lesssim  8$ and
$|C_{10}| \lesssim 100$,  ($|\tilde C_9|< 10^{-3}$ and $|\tilde
C_{10}| < 0.014$), and induces negligible effects in $D \to \pi \mu ^+ \mu^-$  and $D \to
\mu^+ \mu^-$ decays.

\section{Lepton flavor universality violation}
Lepton flavor universality was checked in the case of
$B \to K \ell ^+ \ell^-$ with $\ell = e, \mu$ by the LHCb
experiment~\cite{Aaij:2014ora} in the low dilepton invariant mass
region, $q^2 \in [1,6] \e{GeV}^2$.  The disagreement between the
measurement and the value predicted within the SM is 2.6
$\sigma$~\cite{Hiller:2014yaa}.  This disagreement might be result of
NP as first pointed out in Ref.~\cite{Hiller:2014yaa}. Many subsequent
studies found a number of models which can account for the observed
discrepancy. In the following we assume that the amplitude for
$D^+ \to \pi^+ e^+ e^-$ receives SM contributions only, while in the
case of $\pi^+ \mu^+ \mu^-$ mode, there can be NP contributions, similarly
to what was assumed for $R_K$ in Ref.~\cite{Becirevic:2015asa}. We
define LFU ratios in the low- and high-$q^2$ regions as
\begin{equation}
R_\pi^\mrm{I} =  \frac{ \BR (D^+ \to \pi^+\mu^+ \mu^-)_{q^2 \in [0.25^2,0.525^2 ] \rm{GeV}^2}}{ \BR (D^+ \to \pi^+ e^+ e^-)_{q^2 \in [ 0.25^2,0.525^2]  \rm{GeV}^2}}\,,
\label{RpiI}
\end{equation}
and 
\begin{equation}
R_\pi^\mrm{II} =  \frac{ \BR (D^+ \to \pi^+\mu^+ \mu^-)_{q^2 \in [1.25^2, 1.73^2] \rm{GeV}^2}}{ \BR (D^+ \to \pi^+ e^+ e^-)_{q^2 \in [1.25^2, 1.73^2 ]  \rm{GeV}^2}  }\,.
\label{RpiI|}
\end{equation}
In the SM the departure of the above ratios from $1$ comes entirely from lepton mass
differences. We find $R_\pi^\mrm{I,SM} = 0.87 \pm 0.09$ in the low-$q^2$ and
$R_\pi^\mrm{II,SM} = 0.999\pm 0.001$ in the high-$q^2$ region, where
in the latter region both leptons are effectively massless.
In Tab.~\ref{tab:LFU} we quote ranges for the ratio $R_\pi^\mrm{II}$ for
the maximal allowed values of Wilson coefficients by rare charm decays
considered in the previous Sections. Generally we find that with currently
allowed Wilson coefficients and assuming no NP contribution in
electronic modes these ratios could become much larger. The
spread in these predictions is large because of unknown relative
phases in the resonant part of the spectrum, i.e., $\BR(D^+ \to \pi^+ e^+ e^-) \approx \BR(D^+ \to \pi^+ \mu^+
\mu^-) \approx (0.5\textrm{--}5.3)\E{-9}$. Note that large
enhancements are allowed in the scenarios
which are currently constrained by $D^+ \to \pi^+ \mu^+ \mu^-$.
In the low-$q^2$ region the
interference terms in $R_\pi^\mrm{I}$
are even more pronounced since the effect of
nearby $\rho$ resonance is interfering either in positive or in negative
direction, and thus we cannot conclude the sign of deviation from the
SM value of $R_\pi^\mrm{I}$.
\begin{table}[t]
\begin{center}
\begin{tabular}{||c||r@{.}l|r@{--}l||} 
\cline{2-5}
 \multicolumn{1}{c||}{}  & \multicolumn{2}{c|}{$|\tilde C_i|_\mathrm{max}$} &  \multicolumn{2}{c|}{$R_\pi^\mrm{II}$} \\ 
\hline \hline
SM & \multicolumn{2}{c|}{-} & \multicolumn{2}{c||}{$ 0.999\pm 0.001$}  \\
$\tilde C_7$ & \hspace{0.8cm} 1&6 \hspace{0.8cm} & \hspace{0.8cm} $\sim
                                                   6$&100 \hspace{0.8cm}  \\
$\tilde C_9$ & 1&3 &  $\sim
                                                   6$&120  \\
$\tilde C_{10}$ & 0&63  & $\sim$ 3&30 \\
$\tilde C_S$ & 0&05 & $\sim$ 1&2\\
$\tilde C_P$ & 0&05 & $\sim$ 1&2 \\
$\tilde C_T$ & 0&76 & $\sim$ 6&70 \\
$\tilde C_{T5}$ & 0&74& $\sim$ 6&60  \\
$\tilde C_9= \pm \tilde C_{10}$ & 0&63&  $\sim$ 3&60\\
$\tilde C_9' = -\tilde C_{10}'\big|_{\mrm{LQ}(3,2,7/6)}$ & 0&34& $\sim$ 1&20\\
 \hline
\end{tabular}
\end{center}
\caption{The LFU ratio $R_\pi^\mrm{II}$ at high dilepton invariant mass bin
  and maximal value of each Wilson coefficient (applies also
  for the primed coefficients, $\tilde C_i^\prime$). It is assumed
  that NP contributes only to the muonic mode. The SM value of
  $R_\pi^\mrm{II}$ is given in the first row.}
\label{tab:LFU}
\end{table}

\section{Summary and Outlook}
Motivated by the great improvement of bounds on rare charm decays by
the LHCb experiment 
we determine bounds on the effective Wilson coefficients. 
Existing data implies upper bounds on the effective Wilson
coefficients as presented in Tab.\ref{tab:BR}. The strongest
constraints on $C_{10}$, $C_P$, $C_S$ and  $C_{10}^\prime$,
$C_P^\prime$, $C_S^\prime$ are obtained from the bound on the
branching fraction of
$D^0 \to \mu^+ \mu^-$ decay. The nonresonant differential decay width
distribution gives bounds on $C_i$, $ i=7,9,10, S, P, T, T5$  as well
as on the coefficients of the operators of opposite chirality. The
constraints are stricter in the high dilepton invariant mass bin than
in the low dilepton invariant mass
bin, and this statement applies in particular to the
contributions of the scalar and pseudoscalar operators. Forward-backward asymmetry is sensitive to the combination of scalar and
tensor coefficients at high-$q^2$.

Then, we have investigated new physics models in which the effective
operators may be generated. We have found that the presence of a
leptoquark which is either scalar and weak doublet, $(3,2,7/6)$, or
has spin-1 and is a weak singlet, $(3,1,5/3)$, can lead to sizeable contributions to the
Wilson coefficients $C_9^\prime$ and $C_{10}^\prime$. Sensitivity to
the LQ scenarios is similar in high-$q^2$ bin of 
$D^+ \to \pi^+ \mu^+ \mu^-$ and $D^0 \to \mu^+ \mu^-$, while 
$D^0 - \bar D^0$ mixing results in somewhat stronger constraint.
For the Two Higgs doublet model of type III the presence of scalar and
pseudoscalar operators enhances sensitivity in $D^0 \to \mu^+ \mu^-$
and therefore results in small effects in $D^+ \to \pi^+ \mu^+ \mu^-$.
We have also discussed a SM extension by a $Z'$ gauge boson where
tree-level amplitude in $D^0 - \bar D^0$ mixing is a dominant
constraint and leaves no possibility of signals in rare charm decays.

Our study indicates a possibility to check whether lepton flavor universality between
muonic and electronic channels is valid  by means of studying 
ratios of widths of $D^+ \to \pi^+ \ell^+ \ell^-$ at low or high
dilepton invariant mass bins, $R_\pi^\mrm{I,II}$. In the
SM the two ratios are close to 1,
especially in the high-$q^2$ bin. Assuming the electronic decay is 
purely SM-like we find that in the high-$q^2$ bin the ratio
$R_\pi^\mrm{II}$ is in most
cases significantly increased with respect to the SM prediction, while there is no clear preference
between higher and lower values at low-$q^2$ bin ratio
$R_\pi^\mrm{I}$. In the leptoquark models studied in this paper the ratio may be greatly
increased, but slight decrease cannot be excluded, presently due to
unknown interplay of weak phases with the phases of resonant spectrum.
Chances to observe new physics in rare charm decays are possible in models
where the connection to the stringent constraints stemming from $B$
and $K$ flavor physics are hindered. New physics models which fulfill
this condition are main candidates to be exposed experimentally by
future progress in bounding rare charm decays $D \to \pi \mu^+ \mu^-$
and $D^0 \to \mu^+ \mu^-$, as well as by more precise studies of
$D^0 - \bar D^0$ mixing with the potential NP contributions.
Alternatively, experimental tests of lepton flavor universality in rare charm decays might point towards
presence of new physics in the charm sector, which can easily be hidden in the case
of existing experimental observables.

\noindent
\begin{center}
\textbf{NOTE}
\end{center}
While we were finishing this paper another work~\cite{deBoer:2015boa} appeared in which
the authors studied rare charm decays.

\acknowledgements We thank Benoit Viaud for constructive
comments on the first version of this paper. We acknowledge support of the Slovenian Research
Agency. This research was supported in part by the Munich Institute
for Astro- and Particle Physics (MIAPP) of the DFG cluster of
excellence "Origin and Structure of the Universe''.

\bibliography{refs}

\end{document}